\documentclass{ws-p8-50x6-00}
\newcommand{\etal}{{\em et al.}}                
\begin{document}
\title{Probing Baryon Freeze-out Density at the AGS with Proton Correlations}
\author{Sergei Y. Panitkin$^7$}
\address{for the E895 Collaboration}
\author{
N.N.~Ajitanand,$^{12}$ J.~Alexander,$^{12}$ M.~Anderson,$^5$ 
D.~Best,$^1$ F.P.~Brady,$^5$ T.~Case,$^1$ W.~Caskey,$^5$ 
D.~Cebra,$^5$ J.~Chance,$^5$ P.~Chung,$^{12}$
B.~Cole,$^4$ K.~Crowe,$^1$ A.~Das,$^{10}$
J.~Draper,$^5$ M.~Gilkes,$^{11}$ S.~Gushue,$^2$
M.~Heffner,$^5$ A.~Hirsch,$^{11}$
E.~Hjort,$^{11}$ L.~Huo,$^6$ M.~Justice,$^7$
M.~Kaplan,$^3$ D.~Keane,$^7$ J. Kintner,$^8$
J.~Klay,$^5$ D.~Krofcheck,$^9$ R.~Lacey,$^{12}$ M.~Lisa,$^{10}$
H.~Liu,$^7$ Y.~Liu,$^6$ R.~McGrath,$^{12}$ Z.~Milosevich,$^3$, G.~Odyniec,$^1$
D.~Olson,$^1$  C.~Pinkenburg,$^{12}$
N.~Porile,$^{11}$ G.~Rai,$^1$ H.-G.~Ritter,$^1$
J.~Romero,$^5$ R.~Scharenberg,$^{11}$ L.~Schroeder,$^1$
B.~Srivastava,$^{11}$ N.~Stone,$^2$ T.J.M.~Symons,$^1$ 
S.~Wang,$^7$ J.~Whitfield,$^3$ T.~Wienold,$^1$ R. Witt,$^7$ 
L.~Wood,$^5$ X.~Yang,$^4$ W.~Zhang,$^6$ Y.~Zhang$^4$\\
}
\address{
$^1$Lawrence Berkeley National Laboratory, Berkeley, California 94720\\
$^2$Brookhaven National Laboratory, Upton, New York 11973\\
$^3$Carnegie Mellon University, Pittsburgh, Pennsylvania 15213\\
$^4$Columbia University, New York, New York 10027\\
$^5$University of California, Davis, California 95616\\
$^6$Harbin Institute of Technology, Harbin 150001, P.~R.~China \\
$^7$Kent State University, Kent, Ohio 44242\\
$^8$St.~Mary's College of California, Moraga, California 94575\\
$^9$University of Auckland, Auckland, New Zealand \\
$^{10}$The Ohio State University, Columbus, Ohio 43210\\
$^{11}$Purdue University, West Lafayette, Indiana 47907\\
$^{12}$State University of New York, Stony Brook, New York 11794\\
}  
\maketitle
\abstracts{
First measurements of the beam energy dependence of the two proton
correlation function in central Au+Au collisions are performed by
the E895 Collaboration at the BNL AGS. The imaging technique of
Brown-Danielewicz is used in order to extract information about
the proton source at freeze-out. Measured correlation functions and
sources do not exhibit significant changes with beam energy. 
}
\section{Introduction}
The sensitivity of the two-proton correlations to the space-time 
extent of nuclear reactions was first noted by
Koonin~\cite{koonin_77}. 
A characteristic peak in the proton correlation
function is due to the interplay of the attractive strong and
repulsive Coulomb interactions along with effects of quantum
statistics. For simple static chaotic
sources, it has been shown~\cite{koonin_77,lednicky_82,nakai_81,ernst_85}
that the height of the correlation peak approximately scales inversely
with the source volume. Later on it was realized that the height of
the peak is also sensitive to various dynamical properties 
of the emitting sorce such as collective
flow~\cite{gong_91}, space-momentum correlations within the
source in general~\cite{panitkin_brown}, etc.
Because of the complex nature of the two-proton final state
interactions only model-dependent and/or qualitative statements were 
possible in proton correlation analysis. 
Recently, it was shown that one can perform model-independent extractions 
of the {\em entire} source function 
$S(r)$ from two-proton correlations, not just its
radii, using imaging techniques~\cite{dbrown_1,dbrown_2,dbrown_3}. 
Furthermore, one can do this without making any {\em a priori}
assumptions about the source geometry or lifetime, etc.\\
\indent In this paper we present preliminary results of the first
measurement of the beam energy dependence of the 
two-proton  correlation function in the central Au+Au collisions at
2,4,6 and $8A$ GeV performed by the E895 Collaboration at the Brookhaven
National Lab (BNL) Alternating Gradient Synchrotron (AGS).
\section[E895]{Experimental Details and Analysis Procedure}
The goal of E895 is to study multiparticle correlations and particle 
production with Au beams incident on a variety of targets, over a
range of AGS energies. 
More information about the E895 experimental setup can be found
elsewhere~\cite{rai_90,bauer_97}.
Beams  of gold ions ($^{197}$Au) were available at different energies
- 2,4,6 and $8A$ GeV. They were used to bombard targets of different
materials- Be, Cu, Ag and Au. Charged particles produced in the
collision were detected with time projection chamber
(TPC)~\cite{rai_90}, positioned inside the MPS magnet, and
multi-sampling ionization chamber (MUSIC)~\cite{bauer_97} located
downstream from the magnet. For the presented results, only
information from the TPC was used. 
The TPC was capable of detecting and
the tracking software of reconstructing up to several hundreds tracks per 
event. 
Particle identification was performed via simultaneous
measurement of particle momentum and specific ionization in the TPC
gas. 
Momentum resolution within the range of correlation
measurements was better than 3\%.
In order to obtain the two-proton correlation function $C_{2}$
experimentally, the mixed event technique was used. We employ the
following definition of the correlation function
\begin{eqnarray} C_2(q_{inv}) =
\frac{N_{tr}(q_{inv})}{N_{bk}(q_{inv})} \hspace{0.3cm}, \end{eqnarray}
\noindent{where}                                            
\begin{eqnarray}                                                     
q_{inv} = q = \frac{1}{2} \sqrt{-(p_1^{\mu} - p_2^{\mu})^2}              
\end{eqnarray}                                                      
\noindent{is the half relative invariant momentum between the two
identical    
particles with four-momenta $p_1^{\mu}$ and $p_2^{\mu}$. The
quantities  N$_{tr}$ and N$_{bk}$
are the ``true'' and ``background'' two-particle distributions
obtained by selecting particles from the same and different events,     
respectively. Before calculating the correlation function, several
cuts were applied~\cite{panitkin_99_1}. In order to guarantee a reliable particle
identification and high purity of the proton sample, a cut on proton
longitudinal momentum $P_z<800$ MeV/c was used.
Contamination of the identified proton sample by other particles, in
this momentum interval, was estimated to be less than $2\%$.
For the present analysis events were
selected with a multiplicity cut corresponding to the upper $5\%$ of
the inelastic cross section for Au+Au collisions. 
\begin{figure}[ht]
\begin{center}
\epsfig{file=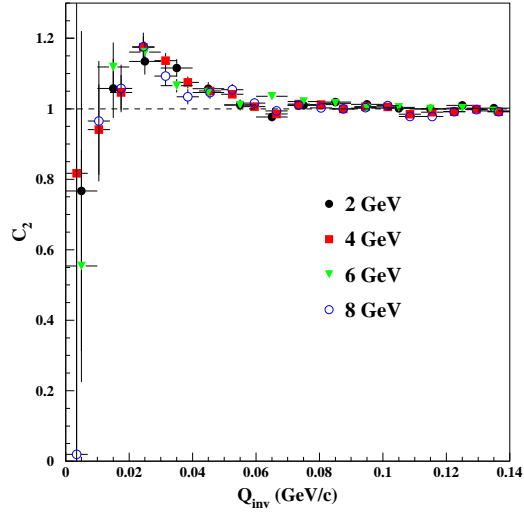, width=3.0in}
\caption{Measured two-proton correlation functions for Au+Au central
collisions at different beam energies. } 
\label{panitkin_1}
\end{center}
\end{figure}
The imaging technique of
Brown-Danielewicz\cite{dbrown_1,dbrown_2,panitkin_brown} was used in
order to extract quantitative information from the measured
correlation functions.  
Here we will give just a brief
sketch of the method, see Refs~\cite{dbrown_1,dbrown_2} for a more detailed description.
The two-particle correlation function may be expressed in the
following way:
\begin{equation}
C_{\bf P}({\bf q}) 
 \simeq
\int d{\bf
r} \, |\Phi_{\bf q}^{(-)}({\bf r})|^2 \, S_{\bf P} ({\bf r}).
\label{panitkin_CPq}
\end{equation}
$S_{\bf P} ({\bf r})$ is the distribution of relative separation 
of emission points for the two particles, in their center of mass and
$\Phi_{{\bf q}}^{(-)}({\bf r})$ is a relative wave function, which can
be calculated numerically given a particular description of the final
state interaction. In the present analysis, the proton relative wave
functions were calculated by solving the  Schr\"odinger equation with
the  REID93\cite{sto94} and Coulomb potentials.
The imaging method takes advantage of the fact that the equation
relating the source and the correlation function may be inverted to yield
a source function $S_p(r)$ from a measured correlation function.  

\section{Results}
Figure~\ref{panitkin_1} shows the measured two-proton correlation
functions for Au+Au central collisions at 2,4,6 and $8A$ GeV. Within
currently available statistical accuracy no significant changes of the
measured correlation functions with beam energy were observed. 
Figure~\ref{panitkin_2} shows the relative distribution of
emission points of protons for central Au+Au collisions at 2,4,6 and
$8A$ GeV obtained as a result of the application of the imaging
technique described above. 
It can be seen from Figure~\ref{panitkin_2} that the relative
proton source functions have similar shapes at all measured
energies.
Imaging combined with the measurements of the single particle
distribution allows one to obtain freeze-out densities in a rather
straightforward way~\cite{dbrown_3}: $\rho = S(r\rightarrow 0)\cdot
N_p$ where $\rho$ is the freeze-out density, $S(r\rightarrow 0)$ is a
value of the source finction at zero separation and $N_p$ is number of
particles(protons). 
Details of the proton single-particle distribution analysis can be found
elsewhere~\cite{witt_1}.
Results of the calculation of proton freeze-out density  are shown (as
a fraction of normal nuclear density) in
Figure~\ref{panitkin_3}. Within experimental errors, proton freeze-out
density in the measured momentum window is fairly constant with a
possible hint of reduction at the highest beam energy. Clearly, better
statistical accuracy is desirable to shed more light on the behavior
of this important observable.\\
\begin{figure}[ht]
\begin{center}
\epsfig{file=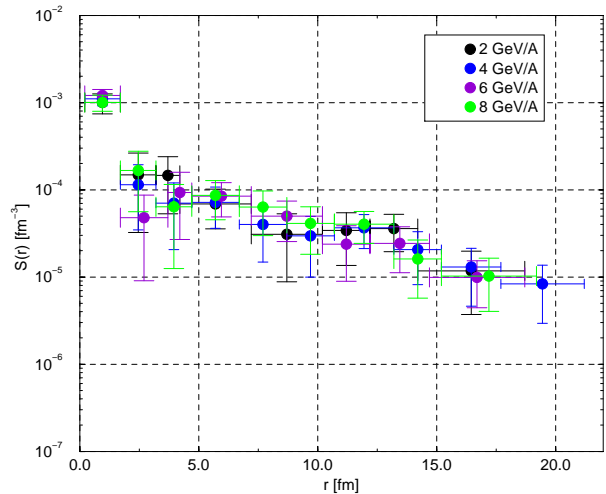, width=2.6 in, angle=-90}
\caption{Relative source functions extracted from the proton
correlation data at 2,4,6 and $8A$ GeV. } 
\label{panitkin_2}
\end{center}
\end{figure}
\begin{figure}[ht]
\vskip4pt
\begin{center}
\epsfig{file=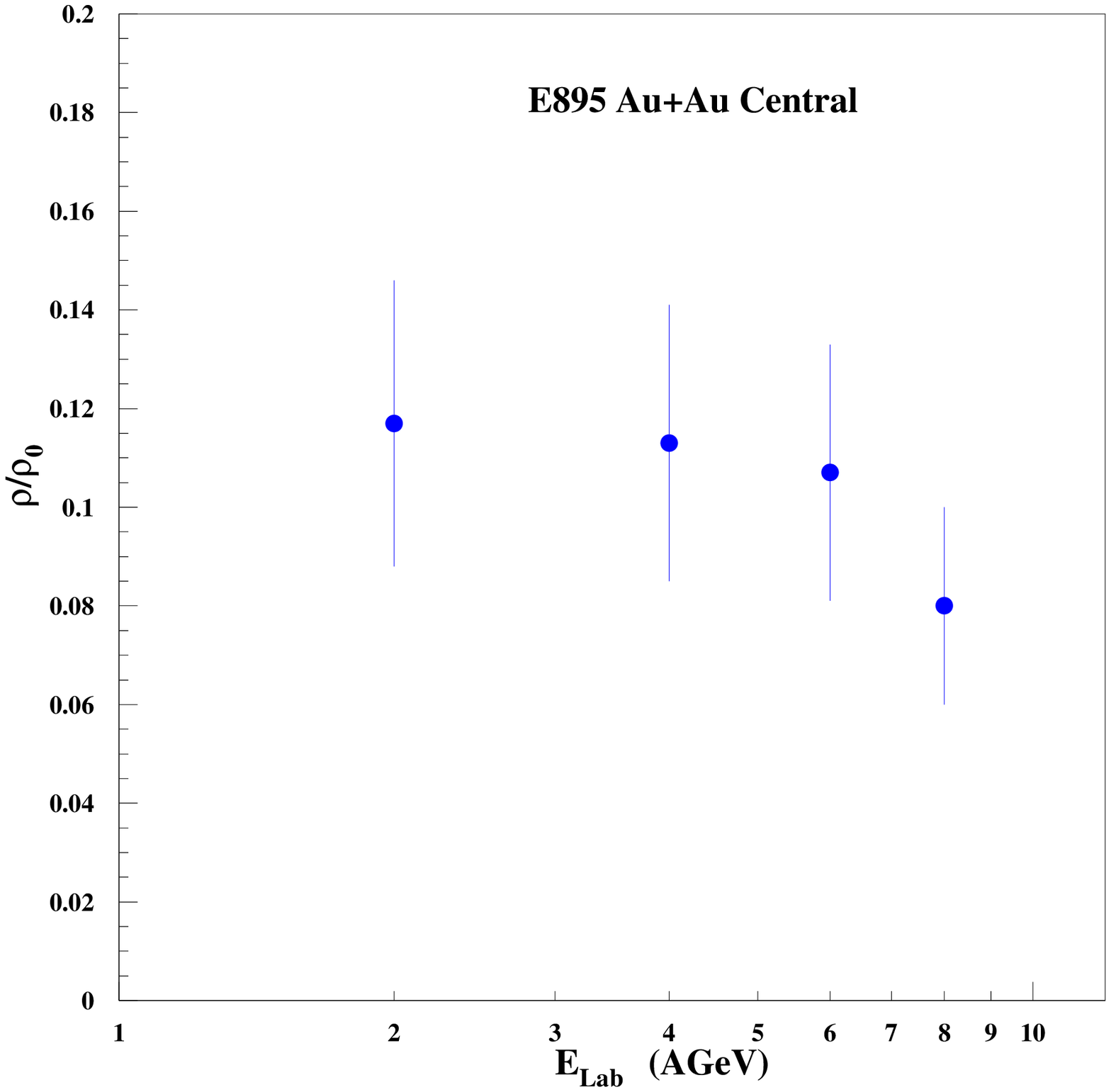, width=3.0in}
\caption{Beam energy dependence of the proton freeze-out densities at
target rapidity region.}  
\label{panitkin_3}
\end{center}
\end{figure}
In summary, we reported preliminary results of the analysis of the
beam energy 
dependence of the two-proton correlation function in the target
fragmentation region ($P_z < 800$ MeV/c). The correlation
functions were measured for the first time for protons in central
Au+Au collisions at beam energies 2,4,6 and $8A$ GeV. Within currently
available statistical accuracy no significant changes with beam energy
were observed. The source imaging technique of Brown-Danielewicz was used
to extract information about the space-time extent of the proton
source. It was found that the relative proton source functions have
similar shapes at all measured energies. 
Measured freeze-out densities of proton's emmited at target hemisphere
($P_z < 800$ MeV/c) were found to be approximately constant at all
available energies. 
Further investigation of this important observable using high
precision measurents is clearly needed.\\
\indent I would like to thank Physics Department of Brown University
for warm hospitality during the Conference. Stimulating discussions
with Drs.~D.Brown and S. Pratt are gratefully acknowledged.  
This research is supported by the U.S. Department of Energy, the
U.S. National Science Foundation and by University of Auckland, New
Zealand, Research Committee. 

\end{document}